\documentclass[a4]{aa}
\usepackage{graphicx}
\usepackage{txfonts}
\usepackage[latin1]{inputenc}

\newcommand{\kms}   {~km~s$^{-1}$}

\newcommand{\jy}    {~Jy~beam$^{-1}$}

\newcommand{\cmd}   {~cm$^{-2}$}
\newcommand{\cmt}   {~cm$^{-3}$}
\newcommand{\vlsr}  {$v_{\rm LSR}$}

\newcommand{\mo}    {$M_{\sun}$}

\newcommand{\T}[1]  {T_{\rm #1}}
\newcommand{\et}    {et al.}
\newcommand{\eg}    {e.\,g.,}

\newcommand{\jr}[2] {\mbox{$J$=#1$\rightarrow$#2}}
\newcommand{\hco}   {HCO$^+$}
\newcommand{\htco}  {H$^{13}$CO$^+$}
\newcommand{\cs}    {C$^{34}$S}
\newcommand{\J}[2]  {\mbox{(#1$\rightarrow#2$)}}

\newcommand{\ie}    {i.\,e.,}
\newcommand{\ar}[3]  {#1$^{\rm h}$#2$^{\rm m}$#3$^{\rm s}$}
\newcommand{\dec}[3] {#1$^{\circ}$#2$'$#3$''$}
\newcommand{\arcdeg}{\mbox{$^\circ$}} 

\begin{document}


\title{Evidence for transient clumps and gas chemical evolution in the CS core
  of L673}
\titlerunning{Transient clumps and gas chemical evolution in L673}

\author{
Oscar Morata\inst{1}
\and Josep Miquel Girart\inst{2,3}
\and Robert Estalella\inst{2}
}
\authorrunning{Morata et al.}

\offprints{Oscar Morata;\\ omorata@mps.ohio-state.edu}

\institute{
Department of Physics, The Ohio State University, 174 West 18th Avenue,
  Columbus, OH 43210
\and
Departament d'Astronomia i Meteorologia, Universitat de Barcelona, Av.\
  Diagonal 647, 08028 Barcelona, Catalunya, Spain 
\and Institut de Ci\`encies de l'Espai (CSIC) / IEEC,
 Gran Capit\`a 2, 08034 Barcelona, Catalunya, Spain
}

\date{Received ...; accepted ...}

\abstract{

  We present FCRAO maps as well as combined BIMA and FCRAO maps of the high
  density molecular emission towards the CS core in the L673 region. With the
  FCRAO telescope we mapped the emission in the CS \J{2}{1}, C$^{34}$S
  \J{2}{1}, HCO$^+$ \J{1}{0}, and H$^{13}$CO$^+$ \J{1}{0} lines.  The
  high-density molecular emission, which arises from a filamentary structure
  oriented in the NW-SE direction, shows clear morphological differences for
  each molecule.
  We find that HCO$^+$ has an extremely high optical depth, and that the
  H$^{13}$CO$^+$ emission is well correlated with submm sources. The BIMA and
  FCRAO combined maps recover emission from structure previously undetected or
  marginally detected, and show an overall aspect of a filamentary structure
  connecting several intense clumps. We found a total 15 clumps in our
  combined data cube, all of them resolved at our angular resolution, with
  diameters in the 0.03--0.09~pc range. Their estimated masses range between
  0.02 and 0.2~\mo, except for the largest clump, which has a mass of
  $\sim1.2$~\mo. We find a clear segregation between the northern and southern
  region of the map: the northern section shows the less chemically evolved
  gas and less massive but more numerous clumps, while the southern region is
  dominated by the largest and most massive clump, and contains the more
  evolved gas, as traced by emission of late-time molecules.  We find that the
  derived clump masses are below the virial mass, and that the clump masses
  become closer to the virial mass when they get bigger and more massive. This
  supports the idea that these clumps must be transient, and that only the
  more massive ones are able to condense into stars. The clumps we detect are
  probably in an earlier evolutionary stage than the ``starless cores''
  reported recently in the literature. Only the most massive one has
  properties similar to a ``starless core''.

}

\maketitle

\authorrunning{Morata et al.}
\titlerunning{Multitransitional observations of L673. II.}

\section{Introduction}
\label{intro}

Dense cores of molecular clouds have been extensively used to study the star
formation process. The emission of several molecules known to be good tracers
of high density molecular gas has been used to study their properties, such as
mass and density, temperature, and velocity distributions.  However, it was
apparent from the first surveys (Zhou et al. \cite{zhou}; Myers et al.
\cite{myers}) that there existed large discrepancies between the emission of
some of these molecules, such as CS, NH$_3$ or HCO$^+$, which needed to be
clarified in order to obtain the actual distribution of the high density gas
in these sources. Pastor et al. (\cite{Pastor91}) and Morata et al.
(\cite{Morata97}) carried out a systematic comparison between the emission of
the CS \mbox{($J$=1$\rightarrow$0)} and NH$_3$ \mbox{($J,K$)=(1,1)}
transitions, under similar conditions of angular resolution, which confirmed
these discrepancies, in particular the systematic difference in size and
separation between the emission peaks of both molecules.

To explain these results, Taylor et al. (\cite{Taylor96}, \cite{Taylor98})
developed a chemical model in which high density condensations, or dense
cores, are formed by clumps of different masses, ages, sizes, and densities
but with sizes of $\la0.1$~pc. These clumps would be unresolved in single-dish
angular resolution observations such as the ones listed in the previous
paragraph.  Most of the clumps would disperse before NH$_3$ abundances build
up to significant levels, though these clumps would contain substantial CS, so
its emission should be observable. A few clumps, those sufficiently
long-lived, or in a more advanced stage of physical and chemical evolution
because of being denser or more massive, would form a significant amount of
NH$_3$, while the CS abundance decreases with time. These clumps would
possibly continue their evolution to eventually form stars. Moreover, Taylor
et al.  (\cite{Taylor96}, \cite{Taylor98}) found that a classification could
be made in the model between {\it ``early-time''} molecules (showing extended
emission like CS) and {\it ``late-time''} molecules (showing more compact
emission like NH$_3$), according to the time at which these molecules reached
their peak fractional abundance.

In order to test the Taylor et al.  chemical model, and in particular to study
the small--size structure of the dense cores, we carried out high angular
resolution observations with the BIMA telescope towards the L673 molecular
cloud of several transitions corresponding to molecules of both families of
species (Morata, Girart \& Estalella \cite{Morata03}; hereafter MGE03).  The
region selected had been previously mapped at an angular resolution of $\sim
1'$ in CS \J{2}{1} (Morata et al. \cite{Morata97}) and NH$_3$
\mbox{($J,K$)=(1,1)} (Sep\'ulveda et al. in preparation). It has no signs
of star formation and is located $\sim9\farcm5$ to the southeast of
IRAS~19180$+$1116.  The distance to the L673 cloud is $\sim300$ pc (Herbig \&
Jones \cite{Herbig83})

MGE03 detect emission in the CS~(\jr{2}{1}), N$_2$H$^+$ (\jr{1}{0}),
\hco~(\jr{1}{0}) and SO ($J_K=2_3\rightarrow1_2$) lines. The high angular
resolution interferometric observations reveal a much clumpier medium than the
lower resolution single-dish observations. Several clumps of size $\la0.08$ pc
are detected distributed all over the region encompassed by the single-dish
observations.  However, the BIMA interferometer only detects 9--12\% of the CS
(2$\rightarrow$1) emission. Modelling the filtering effect of BIMA, MGE03 show
that a clumpy and heterogeneous medium could explain this effect.

MGE03 find that the BIMA observations can be fitted by a model with a density
at which collapse is halted of $n_f=5\times10^5$~cm$^{-3}$. A tentative
classification of the studied clumps according to the stage of chemical
evolution indicated by the molecular abundances is proposed. However, the
model cannot explain the observed low \hco\ abundances with respect to the
predicted values obtained at two positions, but this is possibly due to strong
self-absorption in the \hco\ \J{1}{0} line.

The results of MGE03 show that although the interferometric observations lose
information on the total emission of the region, they are in agreement with
the results obtained with single-dish telescopes, suggesting that
interferometric observations are useful for studying with high angular
resolution the chemical evolution of the inner structures of starless cores.
However, in order to avoid the filtering effect of interferometers, array
observations should be combined with single dish observations. This would give
high resolution maps of the full emission.

In this paper we present the study of the spatial distribution of the
molecular emission in L673 from the maps obtained with the 14-m FCRAO
telescope, and from the maps obtained by combining the BIMA data of MGE03 and
the FCRAO maps presented in this paper.  In Sect.~2 we describe the FCRAO
observations and the BIMA and FCRAO combination procedure. In Sect.~3 we
describe the morphological properties of the molecular emission. In Sect.~4 we
analyze the properties of the molecular cloud observed.

\section{Observations}

\subsection{FCRAO Observations}

     \begin{table}[t]
     \caption[]{Lines Observed with the 14-m FCRAO telescope}
     \label{tfcrao}
     \[
\begin{tabular}
{l@{\hspace{0.1cm}}c@{\hspace{0.1cm}}r@{\hspace{0.1cm}}c@{\hspace{0.1cm}}c@{\hspace{0.1cm}}c}
     \noalign{\smallskip}
     \hline
     \noalign{\smallskip}   
\multicolumn{2}{c}{} &
\multicolumn{1}{c}{$\!\!\!\!\nu$} &
\multicolumn{1}{c}{$\!\!\!\!$FWHM} &
\multicolumn{1}{c}{$\!\!\!\!\Delta v$} & 
\multicolumn{1}{c}{$\!\!\!\!\Delta \T{mb}$} 
\\
\multicolumn{1}{c}{Molecule} &
\multicolumn{1}{c}{Transition} &
\multicolumn{1}{c}{$\!\!\!\!$(GHz)} &
\multicolumn{1}{c}{$\!\!\!\!$Beam} &
\multicolumn{1}{c}{(km$/$s)} &
\multicolumn{1}{c}{$\!\!\!\!\!\!$(K)} 
\\
     \noalign{\smallskip}
     \hline
     \noalign{\smallskip}     
\htco   & \J{1}{0}    & 86.75434 & 59$''$ & 0.17 & 0.04 \\
\hco    & \J{1}{0}    & 89.18852 & 58$''$ & 0.16 & 0.11 \\
\cs     & \J{2}{1}    & 96.41295 & 53$''$ & 0.15 & 0.07 \\
CS      & \J{2}{1}    & 97.98097 & 52$''$ & 0.15 & 0.14 \\
     \noalign{\smallskip}
     \hline
     \end{tabular}
     \]
    \end{table}
%

The observations were carried out in 2003 January with the 14--m Five College
Radio Astronomy Ob\-ser\-vatory (FCRAO) telescope using the SEQUOIA (Second
Quabbin Observatory Imaging Array) 16 beam array receiver.  We used the dual
channel correlator (DCC), which provided a bandwidth of 25 MHz and a spectral
resolution of 24.4~kHz. The DCC was configured to observe the 4 molecular
transitions using 3 different frequency setups: one for \hco, one for CS, and
one for \htco\ and \cs, which were observed simultaneously.  The main-beam
efficiency at the frequencies observed is $\eta_{\rm mb} \simeq 0.50$. At 89
and 98~GHz the half-power beam width is $58''$ and $52''$, respectively. The
weather conditions during the observations were excellent, providing averaged
system temperatures of $\sim$120~K. The observations were performed using the
On-The-Fly (OTF) mapping technique, covering a region of $8'\times 8'$ for
\hco\ and CS and of $6'\times 6'$ for \htco\ and \cs. The center of the maps
was $\alpha (J2000) = 19^{\rm h}20^{\rm m}52\fs20$; $\delta (J2000) =
11\arcdeg14'45\farcs5$.  The selected reference position was 30$'$ east of the
center position. A summary of the observed molecular line transitions is given
in Table~\ref{tfcrao}. The OTF maps were regridded to have a cell size of
20$''$. The line intensities are given in main-beam brightness temperature,
$T_{\rm mb} = T^*_A/ \eta_{\rm mb}$.

\subsection{Combination procedure for the FCRAO and BIMA data}

The BIMA interferometer detects only 9--12\% of the CS \J{2}{1} emission.
Therefore, in order to recover the missing flux of the interferometer data we
combined the CS \J{2}{1}\ and \hco\ \J{1}{0}\ data obtained with the FCRAO
presented in this paper with the BIMA data from MGE03.  The first step to do
this is to convert the FCRAO antenna temperature to a flux scale. We used a
43~Jy~K$^{-1}$ conversion factor for the two transitions (M. Heyer, private
communication).  There are basically two single-dish and interferometric data
combination techniques that work well (Helfer \et\ \cite{Helfer03}). One is to
add synthetic short-spacing visibilities in the $(u,v)$ plane (\eg\ Vogel \et\
\cite{Vogel84}; Girart \et\ \cite{Girart99}). The other is a linear
combination procedure of the single-dish and interferometric maps (\eg\ Ye,
Turtle \& Kennicutt \cite{Ye91}; Stanimirovic \et\ \cite{Stanimirovic99}).  We
tested these two techniques.

In order to fill the BIMA visibility hole with synthetic FCRAO visibilities,
we first deconvolved the FCRAO maps from the FCRAO beam pattern using the
CLEAN algorithm. The beam was modeled as a Gaussian with the FWHM given in
Table~\ref{tfcrao}.  The deconvolved FCRAO map was then multiplied by the
interferometer primary-beam response, taken as a Gaussian, and centered on the
two position observed with BIMA (see MGE03). From these two resulting maps, a
randomly distributed sample of visibilities was generated in the $(u,v)$ plane
of 0--20~light-ns. The number of visibilities generated was chosen to have a
visibility density similar to the BIMA data for the inner part of the
$(u,v)$-plane.  To check the relative flux calibration of the FCRAO data with
the BIMA data, FCRAO visibilities were generated in the range of 21 to
30~light-ns at the same location as the BIMA visibilities. The two data sets
were compared by vector-averaging the visibilities in this range over
concentric annuli. The values obtained from the two instruments agreed to
within 15\%.

     \begin{table}
     \caption[]{Parameters of the FCRAO and BIMA Combined Maps}
     \label{tcomb}
     \[
     \begin{tabular}{lcc@{\hspace{0.1cm}}c@{\hspace{0.1cm}}c}
     \noalign{\smallskip}
     \hline
     \noalign{\smallskip}   
\multicolumn{4}{c}{} &
\multicolumn{1}{c}{$\!\!\!\!$Rms} 
\\
\multicolumn{1}{c}{Molecular} &
\multicolumn{2}{c}{Synthesized Beam} &
\multicolumn{1}{c}{$\!\!\!\!\Delta v$} &
\multicolumn{1}{c}{$\!\!\!\!$Noise} 
\\
\cline{2-3}
\multicolumn{1}{c}{Transition} &
\multicolumn{1}{c}{HPBW} &
\multicolumn{1}{c}{P.A.} &
\multicolumn{1}{c}{$\!\!\!\!$(km$/$s)} &
\multicolumn{1}{c}{$\!\!\!\!\!$(Jy Beam$^{-1}$)} 
\\
     \noalign{\smallskip}
     \hline
     \noalign{\smallskip}     
\hco\ \J{1}{0}& $13\farcs3\times10\farcs3$ & $-16\arcdeg$ & 0.33 & 0.22 \\
CS   \J{2}{1}& $13\farcs1\times10\farcs3$ & $-17\arcdeg$ & 0.30 & 0.25 \\
     \noalign{\smallskip}
     \hline
     \end{tabular}
     \]
    \end{table}
%

     \begin{figure}
      \begin{center}
       \includegraphics[width=74mm]{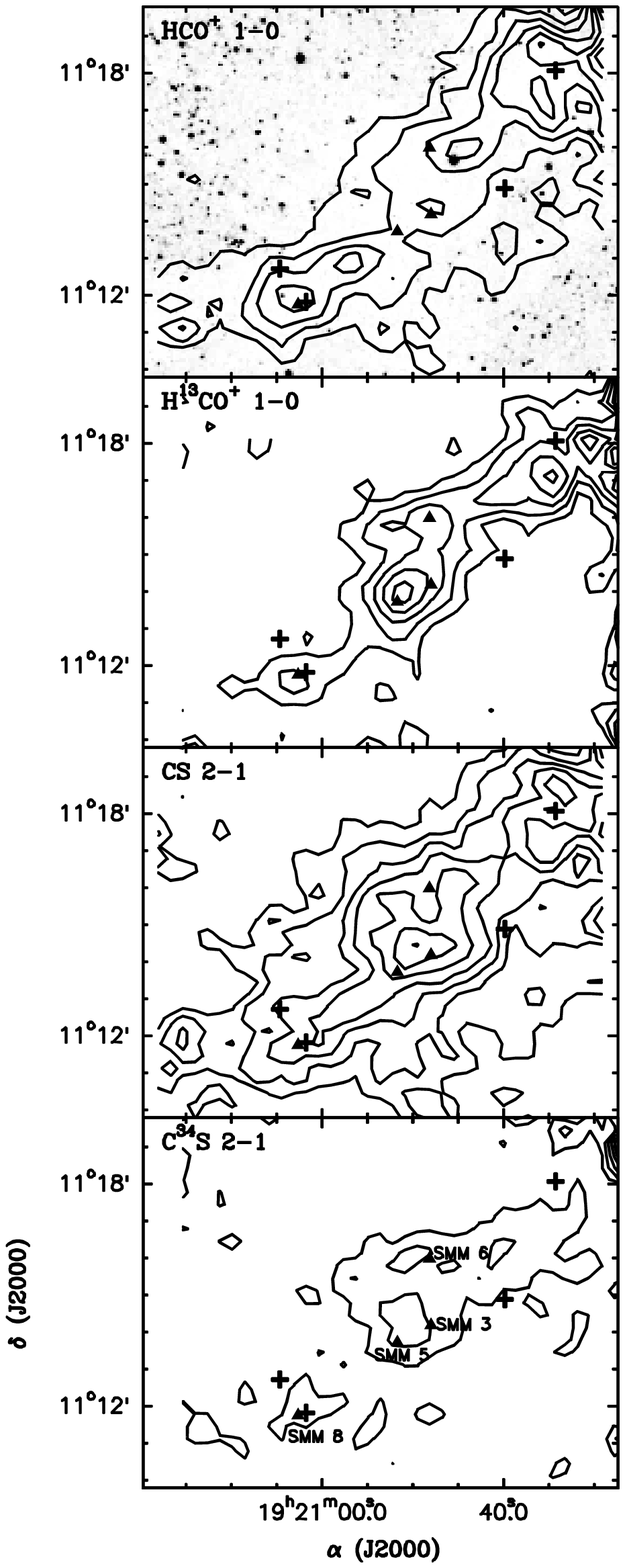}
      \end{center}
      \caption[]{Contour maps of the integrated emission over the \vlsr\ range
      6.05--8.45~\kms\ of (from top to bottom) \hco\ \J{1}{0}, \htco\
      \J{1}{0}, CS \J{2}{1}, and \cs\ \J{2}{1}. The \hco\ contour map is
      overlaid on the J band image from the 2MASS All-Sky Survey. The filled
      triangles show the submm sources reported by Visser, Richer \& Chandler
      \cite{Visser02} (see labels in the bottom panel). The crosses show the
      positions of infrared sources from the IRAS Serendipitous Survey
      Catalog. From west to east: IRAS~19182+1112, IRAS~19183+1109,
      IRAS~19186+1106 and IRAS~19187+1107. Contour levels are 2, 3, 4, \dots\
      8 times 0.24~K~\kms for CS and \hco\ and 1, 2, 3, \dots\ 6 times 0.12
      and 0.086~K~\kms\ for \cs\ and \htco, respectively.  }
      \label{ffcrao} 
     \end{figure}

The linear combination of the maps consists of first generating mosaicked BIMA
dirty maps, $M_{\rm bima}$; as well as the BIMA dirty beam, $B_{\rm bima}$;
then, the "dirty" combined map and beam ($M_{\rm comb}$, $B_{\rm comb}$) are
generated as follows (Stanimirovic \et\ \cite{Stanimirovic99}):
\begin{equation}
M_{\rm comb} = \frac{M_{\rm bima}+f\, \alpha \, M_{\rm fcrao}}{1+\alpha}
\end{equation}
\begin{equation}
B_{\rm comb} = \frac{B_{\rm bima}+\alpha \, B_{\rm fcrao}}{1+\alpha}
\end{equation}
$M_{\rm fcrao}$ is the FCRAO map regridded to match the $M_{\rm bima}$ cell
size and map size and weighted with the primary gain of the mosaicked BIMA
maps; $B_{\rm fcrao}$ is modeled as a Gaussian with the FWHM given in
Table~\ref{tfcrao}; $f$ is the calibration factor, taken as 1.0 given the
agreement of the relative flux calibration between the two instruments; and
$\alpha$ is the ratio of the beam areas for the FCRAO and BIMA. We adopted a
value of $\alpha = 0.050$ and 0.042 for the CS and \hco\ data, respectively.
The different $M_{\rm comb}$ were deconvolved from $B_{\rm comb}$ using the
CLEAN algorithm. Note that modelling the FCRAO beam as a Gaussian is a good
approximation: typical sidelobes in the beam pattern of a single-dish radio
telescope are of the order of $-20$~dB, which will give an error in the
combined beam pattern of less than $10^{-3}$, well below the dynamical range
of the map.

     \begin{figure}
\begin{center}
      \includegraphics[width=7.3cm]{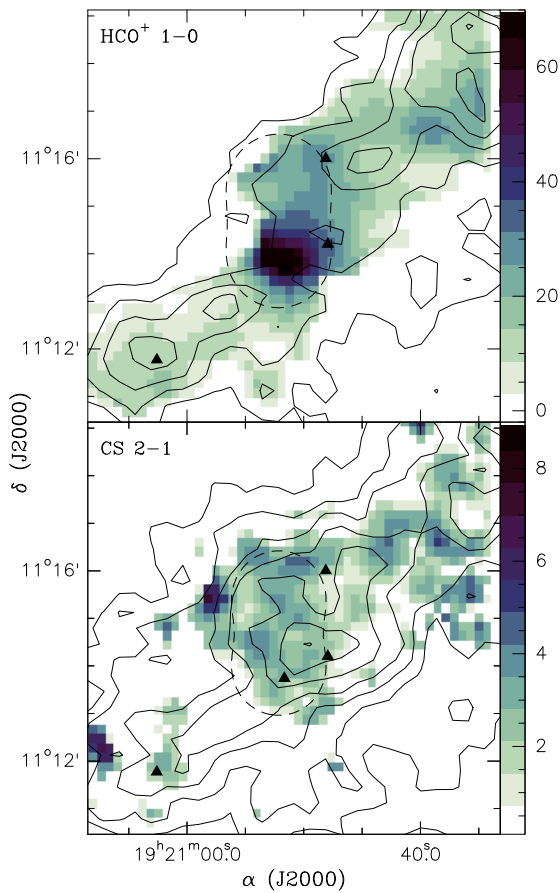}
      \caption[]{Grey scale image of the optical depth of \hco\ \J{1}{0}\ (top
        panel) and CS \J{2}{1}\ (bottom panel) derived from the integrated
        maps of Fig.~\ref{ffcrao}. The scale levels are shown on the right
        side of each panel. The overlaid contour maps show the integrated
        emission of \hco\ and CS. The dashed line shows the gain at the 0.5
        level of the BIMA primary beam response. The symbols are the same as
        those in Fig.~\ref{ffcrao}.}
     \label{ftau} 
\end{center}
     \end{figure}
%

     As pointed out by Helfer \et\ (\cite{Helfer03}), the method of adding
     synthetic single-dish visibilities in the $(u,v)$ plane is
     parameter-dependent, in particular on the number and weighting of the
     synthetic visibilities.  Choosing the right combination of parameters, we
     found results almost identical to those of the linear method technique.
     Yet the linear method is much faster and simpler, so the combined maps
     presented here were obtained using the linear method. Details of the
     combined maps are shown in Table~\ref{tcomb}.

\section{Results and Analysis}

\subsection{Spatial distribution of large scale molecular gas}
\label{large}

Figure~\ref{ffcrao} shows the integrated emission maps of the four species
observed with the FCRAO. The maps show that the high-density molecular
emission arises from a filamentary structure oriented in the NW-SE direction.
Despite sharing the same filamentary structure, the four maps show clear
morphological differences. First, the CS is more spread out than \hco: the
FWHM along the axis perpendicular to the filament major axis is about
$2'$--$3'$ for \hco, whereas is about $3'$--$5'$ for CS.  Second, the \hco\
and \htco\ relative intensity peaks do not coincide. In particular, at the
center of the map they are anti-correlated, \ie\ the \htco\ maximum is located
in a relative minimum of the \hco. Indeed the \htco\ and \hco\ intensities are
comparable at this position. This suggests a very high optical depth. In order
to estimate quantitatively the effect of the opacity on the emission of CS and
\hco\ we estimated their optical depths from the ratio of the integrated
emission maps of \htco\ and \hco\ and from \cs\ and CS, using the isotope
abundance ratios shown in Table~\ref{physparam}. Figure~\ref{ftau} shows the
grey scale image of the \hco\ and CS optical depth overlaid with the contour
map of their integrated emission.  This figure shows clearly that the \hco\ is
highly optically thick, with $\tau \ga 7$ for most of the emitting region and
with a maximum of $\tau = 90$ at the position of L673~SMM~5. On the other
hand, the CS emission is moderately optically thick ($\tau \simeq 1$--$4$).

\begin{table}
\caption[]{Selected positions}
\label{selectedpos}
\begin{tabular}{rlll}
\hline\noalign{\medskip}
id. & \multicolumn{2}{c}{Position} & map counterpart $^{a}$\\ 
\cline{2-3} & $\alpha$ (J2000) & $\delta$ (J2000) & \\
\noalign{\medskip}\hline\noalign{\smallskip}
a  & \ar{19}{20}{34.3} & \dec{11}{18}{05.5}  & IRAS 19182+1112 \\
b  & \ar{19}{20}{43.8} & \dec{11}{16}{05.5}\\
c  & \ar{19}{20}{47.9} & \dec{11}{16}{05.5}  & SMM 6\\
d  & \ar{19}{20}{54.4} & \dec{11}{15}{25.5} & N\\
e  & \ar{19}{20}{54.7} & \dec{11}{14}{45.5}\\
f  & \ar{19}{20}{47.9} & \dec{11}{14}{05.5}  & SMM 3\\
g  & \ar{19}{20}{52.0} & \dec{11}{13}{45.5} & SMM 5 / S\\
h  & \ar{19}{21}{02.9}  & \dec{11}{11}{45.5} & SMM 8 \\
\noalign{\medskip}\hline
\end{tabular}

 \begin{list}{}{}

   \item[$^{a}$] Positions N and S correspond to positions selected for
   study in MGE03.
 \end{list}

\end{table}
%

The extremely high optical depths of the \hco\ \J{1}{0} emission imply that
this line is not a good tracer of the distribution of the \hco\ molecule in
L673.  Thus, we confirm the hypothesis pointed out by MGE03 that the high
optical depth of \hco\ \J{1}{0}\ is responsible for the lack of agreement
between the observed BIMA \hco\ column densities and the theoretically
predicted column densities.  Indeed, the maximum \hco\ optical depth measured
is located within the region observed with BIMA. Strong absorption features
and very high optical depths of the \hco\ \J{1}{0}\ emission are also observed
in other molecular clouds (\eg\ Langer \et\ \cite{Langer78}; Girart \et\
\cite{Girart00}).

\begin{table*}[t]
\caption[]{Physical parameters}
\label{physparam}
\begin{flushleft}
\begin{tabular}{r*{5}{l}cll}
\hline\hline\noalign{\smallskip}
id. & $T_{\rm mb}$ & $T_{\rm mb}$ & $V^{a}$ & $\Delta V^{b}$ & Area & 
$\tau$$^{c}$ &  $N^{d}$\\
 & (CS) & (C$^{34}$S) & (CS) & (CS) & (CS) & (CS) & (CS) \\
    &  (K)  & (K) & (km s$^{-1}$) & (km s$^{-1}$) & (K km s$^{-1}$) & 
    & (10$^{13}$ cm$^{-2}$)\\ 
\noalign{\smallskip}\hline\noalign{\smallskip}
a & 1.34$\pm$0.14 & \hspace{-3.4mm}$<0.14^{e}$ & +7.0 & 0.7 & $1.03\pm0.06$
& \hspace{-3.4mm}$<1.72$ & 0.95$-$1.53$^{f}$ \\
b & 1.64$\pm$0.16 & $0.32\pm0.06$ & +7.2 & 0.8 & $1.48\pm0.06$
& 4.21 & $3.60\pm 0.14$ \\
c & 1.55$\pm$0.17 & $0.26\pm0.04$ & +7.2 & 0.8 & $1.31\pm0.05$
& 3.61 & $2.89\pm0.10$ \\
d & 1.45$\pm$0.18 & $0.24\pm0.06$ & +7.4 & 0.9 & $1.37\pm0.05$
&  3.53 & $2.97\pm0.10$ \\
e & 1.14$\pm$0.13 & $0.19\pm0.06$ & +7.5 & 1.1 & $1.38\pm0.09$
& 3.48 & $2.97\pm0.19$ \\
f & 1.74$\pm$0.21 & $0.32\pm0.05$ & +7.2 & 0.9 & $1.66\pm0.06$
&  3.99 & $3.89\pm0.15$ \\
g & 1.58$\pm$0.27 & $0.35\pm0.06$ & +7.2 & 0.8 & $1.38\pm0.06$
&  5.04 & $3.81\pm0.17$ \\
h & 1.70$\pm$0.16 & $0.19\pm0.06$ & +7.2 & 0.5 & $0.98\pm0.05$
&  2.04 & $1.56\pm0.07$ \\
\noalign{\smallskip}\hline\noalign{\smallskip}
id. & $T_{\rm mb}$& $T_{\rm mb}$ & $V^{a}$ & $\Delta V^{b}$ & Area &
$\tau^{g}$ &  $N^{d}$ & $N$\\ 
 & (HCO$^+$) & (H$^{13}$CO$^+$) & (H$^{13}$CO$^+$) & (H$^{13}$CO$^+$) &
(H$^{13}$CO$^+$) & (H$^{13}$CO$^+$) & (H$^{13}$CO$^+$) & (HCO$^+$) \\
 &  (K) & (K)  & (km s$^{-1}$) & (km s$^{-1}$) & (K km s$^{-1}$) 
&    & (10$^{11}$ cm$^{-2}$) & (10$^{13}$ cm$^{-2}$)\\ 
\noalign{\smallskip}\hline\noalign{\smallskip}
a  & $1.49\pm0.23$ & $0.46\pm0.12$ & +7.0 & 0.4 & $0.22\pm0.03$ &
0.37 & $3.86\pm0.59$ & $2.40\pm0.36$\\
b  & $1.53\pm0.19$ & $0.42\pm0.06$ & +7.1 & 0.5 & $0.23\pm0.03$ &
0.32 & $3.91\pm0.44$ & $2.42\pm0.27$\\
c  & $0.95\pm0.18$ & $0.49\pm0.07$ & +7.2 & 0.7 & $0.37\pm0.02$ &
0.73 & $7.16\pm0.47$ & $4.44\pm0.29$\\
d  & $0.70\pm0.16$ & $0.34\pm0.05$ & +7.4 & 0.7 & $0.25\pm0.03$ &
0.66 & $4.77\pm0.48$ & $2.95\pm0.30$\\
e  & $0.54\pm0.15$ & $0.21\pm0.07$ & +7.5 & 0.5 & $0.12\pm0.02$ &
0.48 & $2.12\pm0.38$ & $1.31\pm0.24$\\
f  & $1.17\pm0.19$ & $0.78\pm0.04$ & +7.3 & 0.5 & $0.38\pm0.02$ &
1.10 & $8.12\pm0.43$ & $5.03\pm0.26$\\
g  & $0.81\pm0.17$ & $0.89\pm0.05$ & +7.3 & 0.5 & $0.45\pm0.02$ &
$>4^{h}$ & \hspace{-3.4mm}$>18.2\pm0.83$ & \hspace{-3.4mm}$>11.3\pm0.51$\\
h  & $2.80\pm0.25$ & $0.42\pm0.97$ & +7.4 & 0.6 & $0.25\pm0.03$ &
0.16 & $4.10\pm0.48$ & $2.55\pm0.30$\\
\noalign{\smallskip}\hline\hline
\end{tabular}

\begin{list}{}{}
  \item[$^{a}$] The error in the line velocity is lower than  0.1\kms. 

  \item[$^{b}$] The error in the line width is lower than 0.1\kms.

  \item[$^{c}$] Using an abundance ratio CS/C$^{34}$S=20 and $T_{\rm
  ex}=4$~K.

  \item[$^{d}$] Beam averaged column density calculated from the formula in
  Table~3 of MGE03. 

  \item[$^{e}$] Upper limit taken as $3\sigma$, where $\sigma$ is the
  sensitivity per channel. 

  \item[$^{f}$] Range of values for $\tau \ll 1$ and $\tau=1.72$.

  \item[$^{g}$] using an abundance ratio HCO$^+$/H$^{13}$CO$^+=62$ (Langer \&
    Penzias~\cite{Langer93}) and $T_{\rm ex}=4$~K.

  \item[$^{h}$] We adopt a value of $\tau=4$ as a lower limit.
\end{list}
\end{flushleft}
\end{table*}

From an inspection to the IRAS catalogues within the region shown in
Fig.~\ref{ffcrao} we found four infrared sources.  IRAS 19182+1112 and
IRAS~19183+1109 are detected only at 12 and 25~$\mu$m. On the other hand, the
other two sources, IRAS~19186+1106 and IRAS~19187+1107 are detected only at
100~$\mu$m. There are also four weak submillimeter sources (Visser, Richer \&
Chandler \cite{Visser02}).  Only the strongest submm source, L673~SMM8, is
associated with an infrared source, IRAS~19187+1107. All the submm sources are
located in strong-molecular-line-emitting regions.  Indeed, they are well
correlated with the relative intensity peaks of \htco\ \J{1}{0}.  Visser
et al. (\cite{Visser02}) classify these submm sources, except SMM3, as
starless cores. SMM3 is classified as a candidate protostar (where there is
some evidence of high-velocity CO emission).

By analyzing the kinematical properties of the CS and \htco\ emission, we found
a smooth velocity gradient along the major axis of the filamentary molecular
structure: from NW to SE the \vlsr\ goes from $\sim 7.0$ to $7.4$~\kms. In
addition, east of the center of the map the CS emission is redshifted to
$\sim7.7$\kms.

\subsubsection{Comparison of the emission of the CS \J{1}{0} and \J{2}{1}
  lines}

We have also compared the CS \J{1}{0} line measured with the Yebes telescope
(Morata \et\ \cite{Morata97}) with the CS \J{2}{1} line obtained with the
FCRAO telescope, after convolving the spectra of the \J{2}{1} line with the
1.9$'$ beam of the Yebes telescope. We compared the resulting spectra at the
position of the CS \J{1}{0} peak, corresponding to the North field of the BIMA
map, and the position of the peak NH$_3$ intensity, corresponding to the South
field. The derived $T_{\rm ex}$ at the North position is the same for both
lines, 4.2~K. In the South position, where we did not observe the C$^{34}$S
\J{1}{0} transition, the excitation temperature derived from the \J{2}{1}
line, 4.4~K, is very similar to the value found in the northern region.

\subsubsection{Column densities}
\label{columndensities}

As in MGE03, we selected several positions in the region mapped with
the FCRAO observations in order to obtain the physical parameters of the gas
and to compare them with the chemical model. Table~\ref{selectedpos} shows the
coordinates for each position and their map counterparts, if any. These
positions were selected mainly to coincide with the 4 positions we selected in
MGE03, and with the SMM and IRAS sources of the region. We also selected two
positions not related to any object in order to see if they traced ``younger''
gas. Table~\ref{physparam} shows the line parameters for the CS \J{2}{1} and
H$^{13}$CO$^+$ lines, and the physical parameters obtained for each
position. We adopted a value of $T_{\rm ex}=4$~K for our calculations, based
on the result we had obtained from the comparison between the CS \J{1}{0} and
\J{2}{1} lines (Morata \et\ \cite{Morata97} and MGE03, respectively).

\begin{table}[t]
\caption{Ratio of column densities}
\label{columnratio}
\begin{tabular}{cc}
\hline\noalign{\smallskip}
id. & $N$(HCO+)/$N$(CS) \\
\noalign{\smallskip}\hline\noalign{\smallskip}
a & \hspace{4mm}1.6--2.5$^{a}$\\
b & $0.67\pm0.08$\\
c & $1.54\pm0.11$\\
d & $0.99\pm0.10$\\
e & $0.44\pm0.08$\\
f & $1.29\pm0.08$\\
g & \hspace{-3.4mm}$>2.96\pm0.19$\\
h & $1.63\pm0.21$\\
\noalign{\smallskip}\hline\noalign{\smallskip}
\end{tabular}

\begin{list}{}{}

  \item[$^{a}$] The value 1.6 is obtained for $\tau=1.72$. The value 2.5 is
  obtained when $\tau\ll1$.

\end{list}

\end{table}

Table~\ref{columnratio} shows the ratio between the derived column densities
for CS and HCO$^+$ obtained for each position, which can be compared to the
results of our chemical model (MGE03). We observe that there is a similar
trend to the results we obtained in MGE03. In the positions where there are no
SMM or IRAS sources (positions b, d, and e) the CS column density is higher
than the HCO$^+$ column density, which would indicate that the clumps are
chemically young. In the positions where there is a SMM or IRAS source, the
column density fo HCO$^+$ is higher than for CS. This is what we expect for
later times, when the clumps are more chemically evolved, HCO$^+$ is more
abundant, and CS starts to suffer freeze-out. We found the highest ratio for
SMM5 (position g), which we previously classified with the BIMA observations
as one of the two chemically more evolved points, and where the N$_2$H$^+$
emission is more intense. Thus, we confirm our previous classification.

\subsubsection{CS abundances}

Visser \et\ (\cite{Visser02}) calculate the masses of the SMM sources in L673
from the total flux measured in an area of radius $r$. They also calculated
the H$_2$ column density over this area. We have convolved the spectra of the
CS \J{2}{1} and C$^{34}$S \J{2}{1} emission with a Gaussian with FWHM equal to
the $r$ derived for each SMM source, in order to estimate the CS column
density for each SMM position. Table~\ref{CSabundance} shows the averaged CS
abundance obtained from the H$_2$ and CS column densities. The derived values
are in the range 2--$10\times10^{-9}$ \cmd. We found the lowest value near the
IRAS source at the SE part of the large scale maps, and the highest abundance
at the position of SMM6, a submm source that seems to be less centrally peaked
than other submm sources (Visser \et\ \cite{Visser02}), which would indicate
that it is in a less evolved stage.

\begin{table}
 \caption[]{Calculated CS abundances at the positions of the SMM sources}
 \label{CSabundance}
 \begin{tabular}{llllllll}
\noalign{\smallskip}\hline\noalign{\smallskip}
  & $\!\!\!\!\!\!\!\!$Beam & $\!\!\!\!\!\!N$(H$_2$)$^{a}$ & & & $N$(CS) &\\
  & $\!\!\!\!\!\!\!\!\!\!\!$(arcsec) & $\!\!\!\!$(cm$^{-2}$) &
  $\!\!\!\!\!T_{\rm C^{34}S}/T_{\rm CS}$ & $\!\!\!\!\!\!\tau$ & (cm$^{-2}$)&
  $\!\!\!\!X$(CS) \\
\noalign{\smallskip}\hline\noalign{\smallskip}
$\!\!\!\!$SMM3 & $\!$60  & $\!\!\!\!\!\!$4.3$\times10^{21}$ &
  $\!\!\!\!$0.15$\pm0.02$ & $\!\!\!\!\!\!$3.2 & $\!\!\!\!$2.74$\times10^{13}$ &
  $\!\!\!\!$6.4$\times10^{-9}$ \\  
$\!\!\!\!$SMM5 & $\!$80  & $\!\!\!\!\!\!$4.0$\times10^{21}$ &
  $\!\!\!\!$0.18$\pm0.02$ & $\!\!\!\!\!\!$3.8 & $\!\!\!\!$2.66$\times10^{13}$
  & $\!\!\!\!$6.7$\times10^{-9}$ \\  
$\!\!\!\!$SMM6 & $\!\!\!\!$100 & $\!\!\!\!\!\!$2.5$\times10^{21}$ &
  $\!\!\!\!$0.15$\pm0.02$ & $\!\!\!\!\!\!$3.2 & $\!\!\!\!$2.56$\times10^{13}$
  & $\!\!\!\!$1.0$\times10^{-8}$ \\  
$\!\!\!\!$SMM8 & $\!\!\!\!$110 & $\!\!\!\!\!\!$3.8$\times10^{21}$ &
  $\!\!\!\!$0.11$\pm0.02$ & $\!\!\!\!\!\!$2.0 & $\!\!\!\!$1.08$\times10^{13}$
  & $\!\!\!\!$2.8$\times10^{-9}$ \\  
\noalign{\smallskip}\hline\noalign{\smallskip}
 \end{tabular}

 \begin{list}{}{}
 \item[$^{a}$ From Visser \et\ \cite{Visser02}]
 \end{list}
\end{table}

\subsection{Spatial distribution at high angular resolution}

\begin{figure*}[t]
      \includegraphics[width=\hsize]{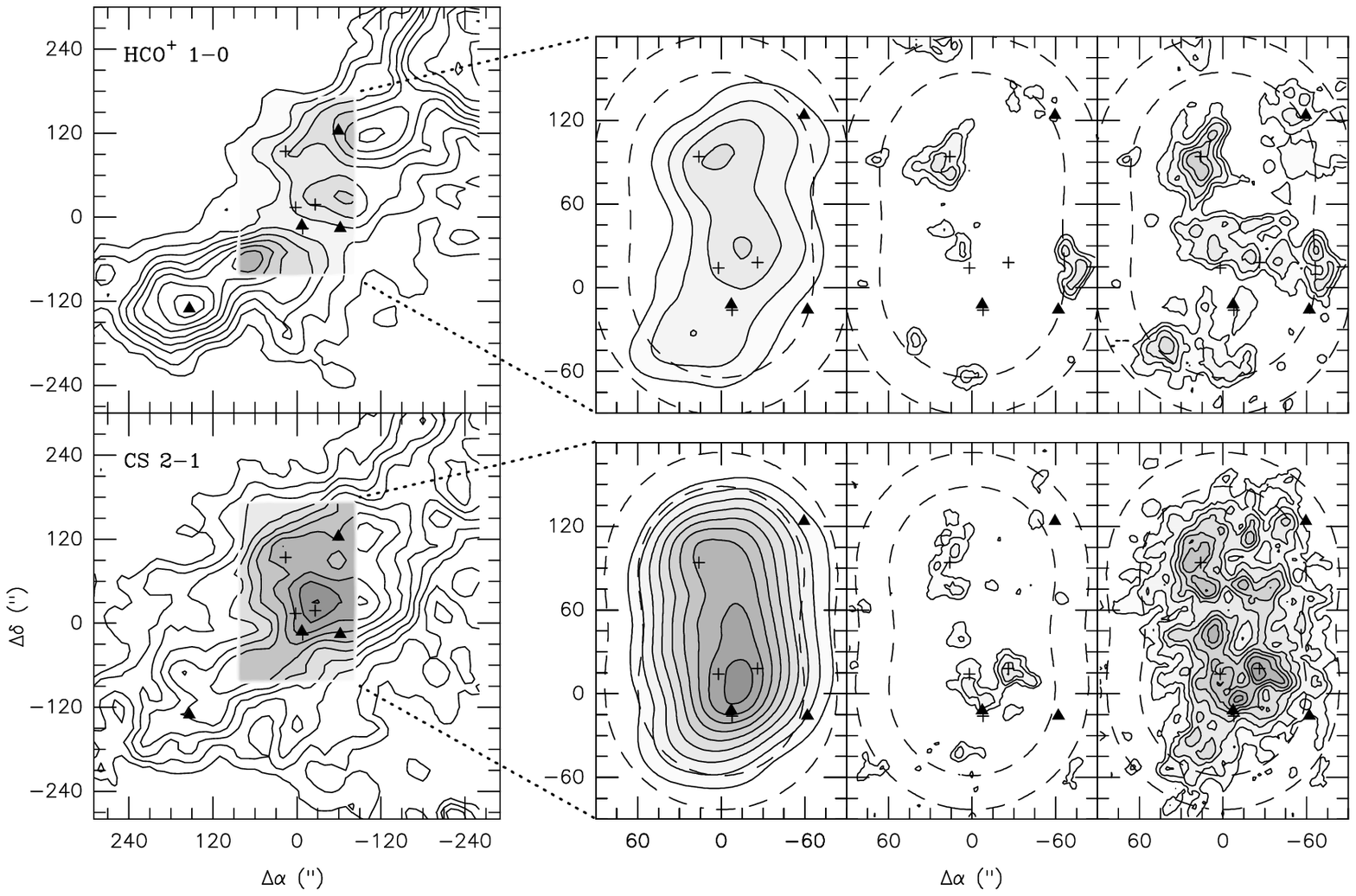}

      \caption[]{ Composite of the combination process for CS \J{2}{1}
      (bottom panels) and \hco\ \J{1}{0} (top panels). All maps show the
      integrated emission over the \vlsr\ range 6.05--8.45~\kms.  Left panels:
      FCRAO maps. Middle-left panels: FCRAO maps corrected for the primary
      beam of the mosaicked BIMA maps. Middle-right panels: mosaicked BIMA
      maps. Right panels: Combined BIMA and FCRAO maps. For the FCRAO maps,
      the contour levels are 3, 4, 5, \dots\ 12 times 3.17 and
      2.80~Jy~beam$^{-1}$~km~s$^{-1}$ for CS \J{2}{1} and \hco\ \J{1}{0},
      respectively. For the BIMA and combined maps, the contour levels are 2,
      3, 4, \dots\ 9 times 0.27 and 0.24~Jy~beam$^{-1}$~km~s$^{-1}$ for CS
      \J{2}{1} and \hco\ \J{1}{0}, respectively. The inner and outer dashed
      lines show the 0.5 and 0.25 level of the BIMA primary beam
      response. The symbols are the same as those in Fig.~1.  }
      \label{fcombine} 
     \end{figure*}

Figure~\ref{fcombine} shows the composite of the combination process of the
BIMA and FCRAO observations for the CS \J{2}{1} and \hco\ \J{1}{0} emission.

%
%

%
     \begin{figure*}[t]
      \includegraphics[width=\hsize]{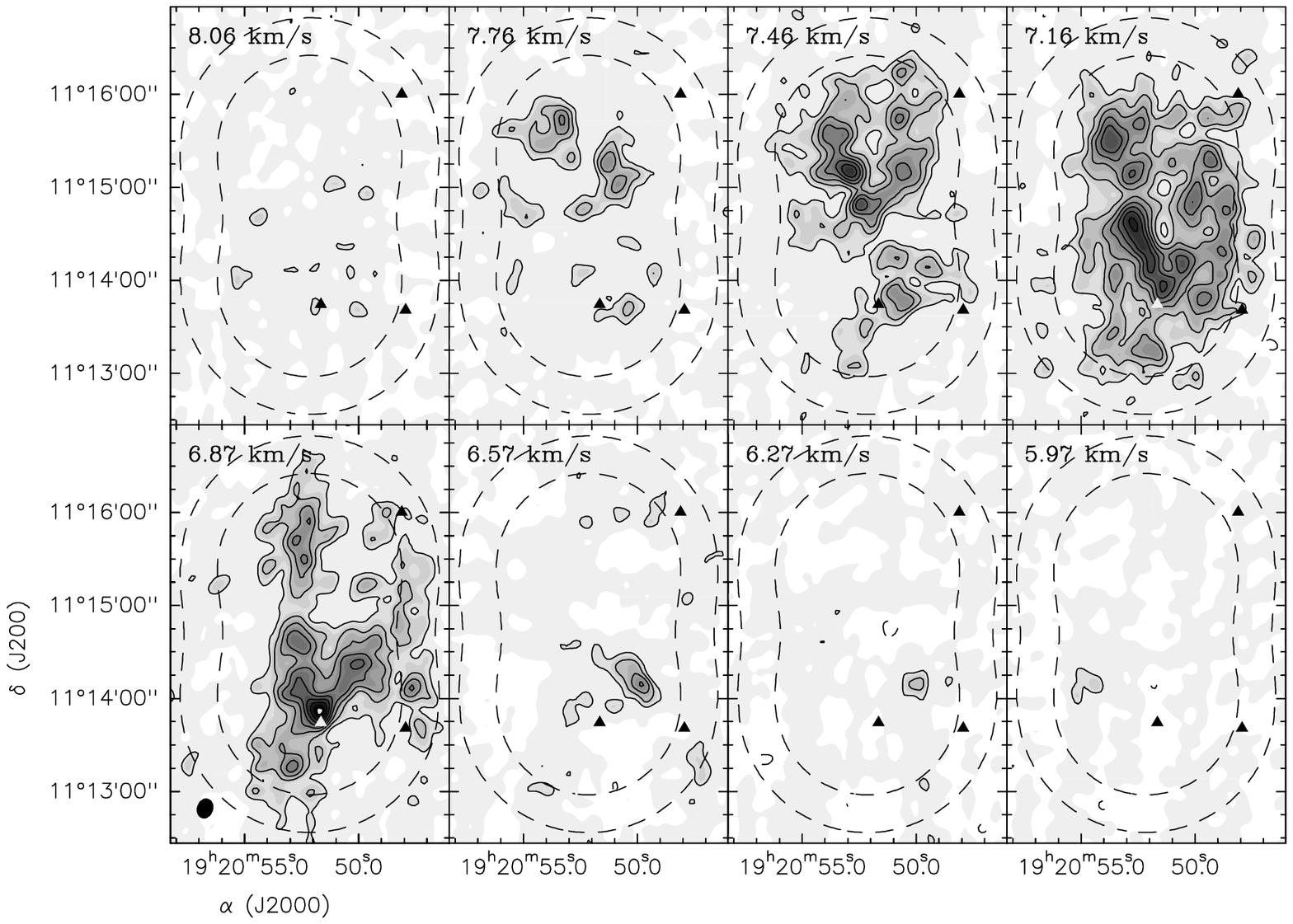}
     \caption[]{     
     Channel maps of the CS \J{2}{1} emission. The
     contour levels are -3, 3, 4.5, 6, \dots\ 12 times 0.24~Jy~beam$^{-1}$
     (the rms noise of the maps). The inner and outer dashed lines show the
     0.5 and 0.25 level of the BIMA primary beam response. The
     symbols are the same as in Fig.~1.
}
     \label{fchannel} 
     \end{figure*}
%

The \hco\ combined map shows a clear filamentary structure connecting several
clumps, although some of them, particularly to the SE, are far from the phase
center of the map, which makes it difficult to distinguish them from another
filament. However, as indicated by Fig.~\ref{ftau}, the region observed has a
very high optical depth, so the emission shown in the combined map cannot be
used to describe the true morphological and kinematical behaviour of the \hco\
emission.

%
\begin{figure*}[t]
  \begin{center}
      \includegraphics[width=\hsize]{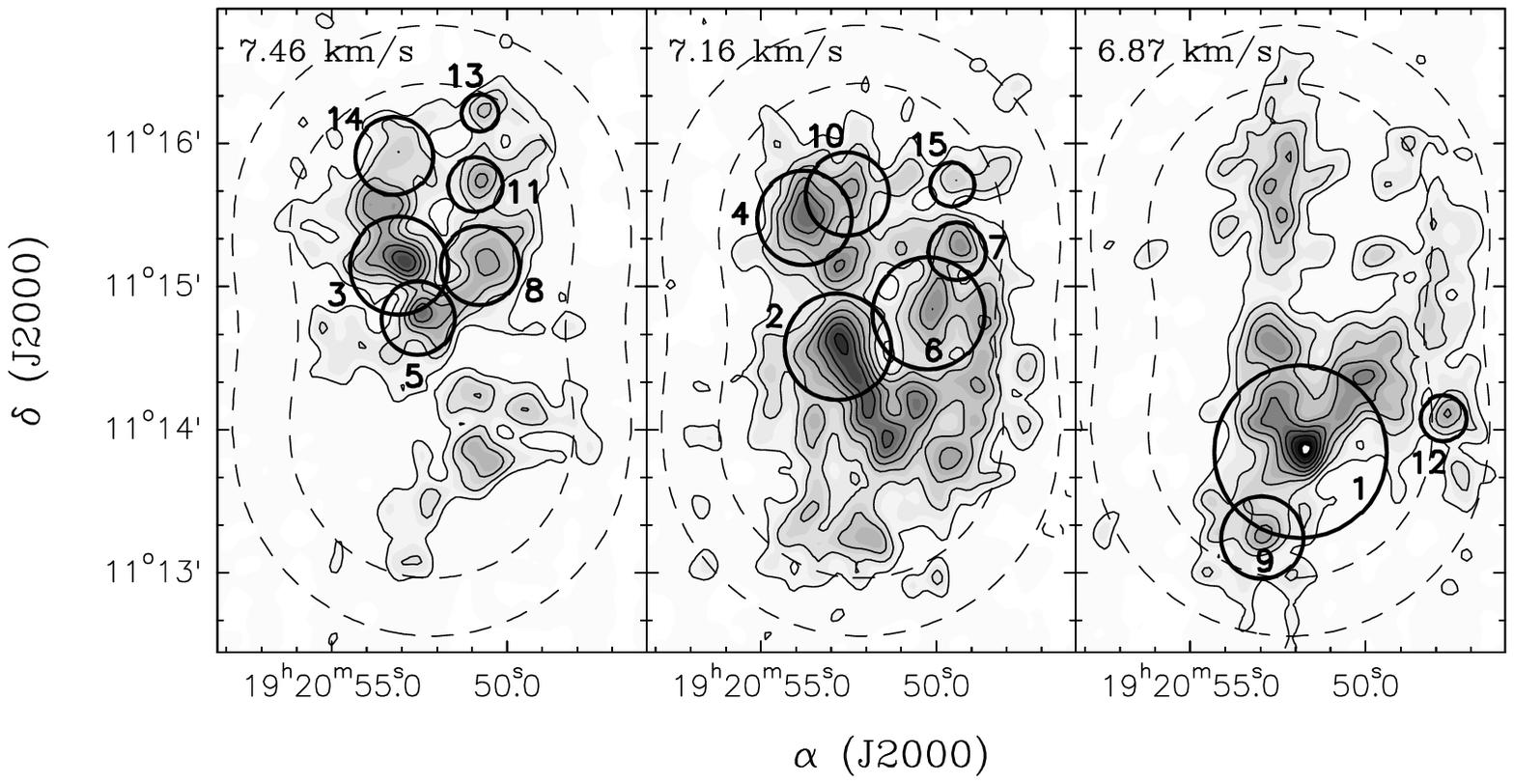}
      \caption[]{ Combined BIMA and FCRAO CS \J{2}{1} contour maps for the
        three channels with the strongest emission (from Fig.~\ref{fchannel}).
        The thick circles are centered at the position and at the channel at
        which each clump shows its highest intensity, as found by the CLFIND
        algorithm of Williams et al. (\cite{Williams94}) and described in
        Sect.~\ref{Sclump}.  The areas of the circles are proportional to the
        masses of the clumps. The numbers correspond to the identifier in the
        first column of Table~\ref{tclumpO}. }
     \label{fclumps} 
     \end{center}
\end{figure*}
%

On the other hand, the combined map of the CS emission, although it has a
moderately high opacity (2--4), can be used to trace the CS distribution.
When we recover all the flux, we find that all the most prominent clumps found
in the BIMA observations are also detected but with more intensity.  We
believe that these clumps, which lie near the phase center of the map, are
real. We also recover emission from a great deal of other structure which was
not, or only marginally, detected previously. This confirms the predicted
effect of the BIMA filtering in MGE03. The overall aspect is also of a
filamentary structure connecting several intense clumps. One of these clumps
coincides with the nominal position of the source SMM~5.

The clumps and the filamentary structure of the emission are more clearly
revealed in Fig.~\ref{fchannel}, which shows the channel maps of the combined
BIMA and FCRAO maps for the CS \J{2}{1}\ line. Here we find a very complex
structure, with several clumps with sizes of $\sim 13''$--$25''$, appearing in
two or three of the channel maps. Some of these clumps might coincide with
clumps found in the HCO$^+$ map.

\subsection{Clump Analysis}
\label{Sclump}

Since the combined high-angular-resolution maps show a high\-ly clumpy CS
spatial and velocity structure, we carried out the clump analysis technique
developed by Williams, de~Geus \& Blitz (\cite{Williams94}) in our combined
BIMA and FCRAO CS \J{2}{1} data cube.  This technique consists mainly of two
algorithms, CLFIND and CLSTATS. CLFIND, which does not assume any clump
profile, works by contouring the data at a multiple of the rms noise of the
data cube, then searches for peaks of emission to locate the clumps, and then
follows them down to lower intensities. CLSTATS is a task that calculates
integrated intensities, peak positions, sizes, velocity dispersion, etc, for
the clumps found by CLFIND. A very detailed explanation of this analysis
technique is found in Williams et al. (\cite{Williams94}), who also found that
the CLFIND algorithm obtains the best results when the contour interval and
the lower contour are set to 2$\sigma$.  However, they recommend using a
higher value of the lower contour if there is background emission.  This is
the case for the combined L673 CS \J{2}{1} maps, where there is some low-level
($\sim2\sigma$) extended emission.  Thus, we run the CLFIND and CLSTATS
algorithms for the L673 CS channel maps with the contour interval set to
0.48~\jy\ (2$\sigma$) and the lowest contour set at 0.96~\jy\ (4$\sigma$).

     \begin{table}
     \caption[]{Clumps within the BIMA primary beam}
     \label{tclumpO}
     \[
     \begin{tabular}{rccccccc}
     \noalign{\smallskip}
     \hline
     \noalign{\smallskip}   
\multicolumn{1}{c}{N} &
\multicolumn{1}{c}{$\!\!\!T_{\rm peak}$} &
\multicolumn{1}{c}{$\!\!\!\tau_{\rm CS}^a$} &
\multicolumn{1}{c}{$\!\!\!\!\Delta v^b$} & 
\multicolumn{1}{c}{$\Delta R^c$} & 
\multicolumn{1}{c}{$\!\!M_{\rm cl}^d$} &
\multicolumn{1}{c}{$\!\!\!M_{\rm vir}^e$} &
\multicolumn{1}{c}{$\!\!\!\!\!n$(H$_{2})^f$}
\\
 & $\!\!\!$(K) & & $\!\!\!\!$(km s$^{-1}$) & (pc) &
     $\!\!$($M_{\odot}$) & $\!\!$($M_{\odot}$) &
     $\!\!\!\!\!\!\!\!(10^4$cm$^{-3}$) \\
     \noalign{\smallskip}
     \hline
     \noalign{\smallskip}
  1 & $\!\!\!$3.5 & $\!\!\!$3.2 & $\!\!\!$0.49 & 0.045 & $\!\!$0.36 & $\!\!$2.28 & $\!\!\!$1.9 \\
  2 & $\!\!\!$3.0 & $\!\!\!$3.2 & $\!\!\!$0.39 & 0.037 & $\!\!$0.14 & $\!\!$1.18 & $\!\!\!$1.3 \\
  3 & $\!\!\!$2.8 & $\!\!\!$2.7 & $\!\!\!$0.43 & 0.023 & $\!\!$0.12 & $\!\!$0.89 & $\!\!\!$4.6 \\
  4 & $\!\!\!$2.5 & $\!\!\!$2.8 & $\!\!\!$0.55 & 0.026 & $\!\!$0.11 & $\!\!$1.65 & $\!\!\!$3.0 \\
  5 & $\!\!\!$2.5 & $\!\!\!$2.7 & $\!\!\!$0.51 & 0.020 & $\!\!$0.07 & $\!\!$1.08 & $\!\!\!$4.1 \\
  6 & $\!\!\!$2.2 & $\!\!\!$3.1 & $\!\!\!$0.38 & 0.036 & $\!\!$0.15 & $\!\!$1.10 & $\!\!\!$1.6 \\
  7 & $\!\!\!$2.1 & $\!\!\!$2.6 & $\!\!\!$0.37 & 0.025 & $\!\!$0.04 & $\!\!$0.73 & $\!\!\!$1.2 \\
  8 & $\!\!\!$2.0 & $\!\!\!$3.5 & $\!\!\!$0.49 & 0.022 & $\!\!$0.08 & $\!\!$1.12 & $\!\!\!$3.4 \\
  9 & $\!\!\!$2.0 & $\!\!\!$2.4 & $\!\!\!$0.31 & 0.033 & $\!\!$0.08 & $\!\!$0.64 & $\!\!\!$1.2 \\
 10 & $\!\!\!$2.0 & $\!\!\!$3.0 & $\!\!\!$0.56 & 0.021 & $\!\!$0.09 & $\!\!$1.39 & $\!\!\!$4.4 \\
 11 & $\!\!\!$1.9 & $\!\!\!$2.7 & $\!\!\!$0.44 & 0.020 & $\!\!$0.04 & $\!\!$0.80 & $\!\!\!$2.3 \\
 12 & $\!\!\!$1.9 & $\!\!\!$1.9 & $\!\!\!$0.57 & 0.027 & $\!\!$0.03 & $\!\!$1.83 & $\!\!\!$0.6 \\
 13 & $\!\!\!$1.6 & $\!\!\!$4.0 & $\!\!\!$0.39 & 0.015 & $\!\!$0.02 & $\!\!$0.47 & $\!\!\!$2.5 \\
 14 & $\!\!\!$1.4 & $\!\!\!$3.0 & $\!\!\!$0.65 & 0.022 & $\!\!$0.07 & $\!\!$1.98 & $\!\!\!$3.3 \\
 15 & $\!\!\!$1.4 & $\!\!\!$2.0 & $\!\!\!$0.39 & 0.021 & $\!\!$0.02 & $\!\!$0.67 & $\!\!\!$1.2 \\
     \noalign{\smallskip}
     \hline
     \end{tabular}
     \]
     \begin{list}{}{}
     \item[$^{a}$] Optical depth derived at the peak position of the clump from
     the  intensity ratio of the CS \J{2}{1} and C$^{34}$S \J{2}{1} FCRAO maps.
     \item[$^{b}$] FWHM of the line corrected for channel resolution as
       described in Appendix A of Williams et al. (1994)
     \item[$^{c}$] Equivalent circular radius corrected for beam size as 
     described in  Appendix A of Williams et al. (1994)
     \item[$^{d}$] Clump mass derived from the integrated emission 
     obtained with CLSTATS corrected for the CS \J{2}{1} optical depth,
     $\tau$,  and assuming a constant excitation temperature of
     $T_{ex}=4$~K. An abundance  ratio of $X[{\rm CS/H}_2]=10^{-8}$ is
     adopted.
     \item[$^{e}$] Virial mass obtained from 
     $M_{\rm vir}=5 \, \Delta R \, \sigma_v^2/\alpha \, G $ with 
     $\sigma_v=\Delta v/2.355$ and assuming an inverse-square power law
     density  profile ($\alpha$=5/3).
     \item[$^{f}$] Average volume density obtained assuming a sphere of mass
     $M_{\rm cl}$ and radius $\Delta R$. 
     \end{list}
\end{table}
%

A total of 15 clumps were found, all of them with a peak intensity of at least
$\sim 4\sigma$.  Table~\ref{tclumpO} lists the physical parameters derived for
the 15 clumps found by the algorithm.  All the clumps are clearly resolved at
our angular resolution, and have deconvolved diameters in the 0.03--0.09 pc
range.  The velocity dispersion of the clumps (\ie\ the FWHM corrected for the
spectral resolution) is not very different for the different clumps,
0.31--0.65~\kms.  The clumps have masses between 0.02 and 0.36~\mo.  The
estimated clump densities range between 0.6 and $4.6\times10^4$~\cmt. This
value of the averaged density of the clumps implies that, since the CS
\J{2}{1} critical density is $5\times10^5$~\cmt, they are subthermalized. It
also explains why the derived excitation temperature, 4 K, is significantly
below the kinetic temperature expected in the cloud, 10 K.

Figure~\ref{fclumps} shows the location of the clumps derived by the algorithm
as circles.  The areas of the circles are proportional to the masses of the
clumps.  This figure shows a segregation of clumps between the northern and
southern section of the map. The clumps with peak intensities in the 7.2 and
7.5~\kms\ channels arise from the northern region, whereas the clumps that
have a peak in the 6.9~\kms\ channel are located in the southern region.  In
addition, the southern part contains fewer clumps than the northern part, but
it is dominated by the largest and most massive clump.  Interestingly, this
clump is associated with the two N$_{2}$H$^+$ clumps found by MGE03.  This
confirms the idea that both N$_{2}$H$^+$ and NH$_{3}$ trace the more evolved,
massive clumps.

\section{Discussion}

\subsection{The large-scale structure of L673. Correlation of HCO$^+$ and N$_2$H$^+$ emission with dust}



As we have pointed out, the \htco\ seems to be well correlated with the dust
as detected by the submillimeter observations. This suggests that \htco\
traces the more massive or more evolved clumps. At the same time,
Fig~\ref{ffcrao} shows that the peak \htco\ emission is also coincident with
the position of the BIMA peak of N$_2$H$^+$, and a SMM source (SMM 5). In our
chemical models, HCO$^+$ and N$_2$H$^+$ are late-time molecules. Thus, the
apparent discrepancy found in MGE03 between the column densities observed with
BIMA and the expected values from the chemical model was due to the extremely
high opa\-city of the HCO$^+$ emission.
\htco\ is found where N$_2$H$^+$ is also found, although a little more
extended, as can be expected from the models, because it is formed a little
earlier. Moreover, the emission of H$^{13}$CO$^+$, a late-time molecule,
coincides with the position of the dusty submm sources, which also agrees with
the theoretical predictions (Taylor \et\ \cite{Taylor98}).

\subsection{The small-scale structure of L673}

\subsubsection{CS clump distribution properties} 

\begin{figure}
      \includegraphics[width=\hsize]{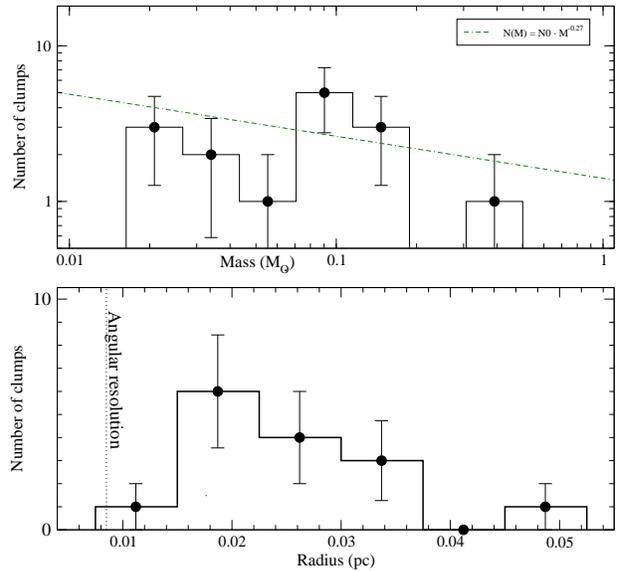}
     \caption[]{     
     {\it Top} Mass distribution of the L673 CS clumps within the primary beam
     of the BIMA  maps. The lines show the clump mass spectrum for a power-law
     distribution, $N \propto M^{-0.27}$.
{\it Bottom} Size (equivalent circular radius) distribution of the L673 CS
clumps.  The dotted line shows the radius of the CS \J{2}{1} synthesized beam.
Note that all the clumps are well resolved at our angular resolution.  }
     \label{fdistri} 
     \end{figure}

Figure~\ref{fdistri} shows the mass and size distribution of the derived
clumps. It is clear that with only 15 clumps we cannot pursue a statistical
analysis of the mass distribution of the clumps. However, with the data
available, the mass clump distribution is compatible with the low-mass end of
the mass distribution observed in low-mass star formation clouds, such as in
$\rho$ Oph (Motte \et\ \cite{Motte98}; Johnstone \et\ \cite{Johnstone00}),
where the slope of $dN/dM$ is not so steep (see Fig.\ \ref{fdistri}).

\begin{figure} [t]
      \includegraphics[width=\hsize]{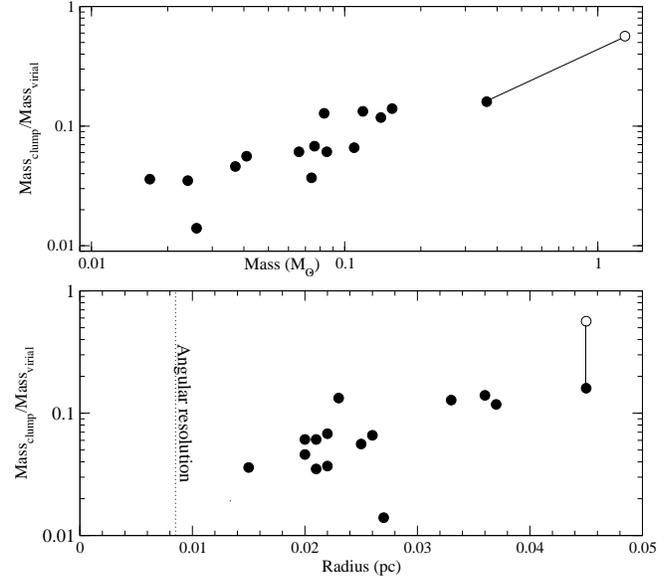}
     \caption[]{     
    Relation between the clump to virial mass ratio with respect to the clump
     mass ({\it top}) and the equivalent circular radius ({\it bottom}). The
    open circle represents the most massive clump when the mass derived from
    the N$_2$H$^+$ emission is considered.
}
     \label{fmass} 
\end{figure}

The derived clump masses are significantly below the virial mass (see
Table~\ref{tclumpO}). This strongly supports the idea that these clumps must
be transient (Taylor et al. \cite{Taylor96}, \cite{Taylor98}).  Recently, a
detailed study of the illuminated clumps ahead of the HH~2 object also
suggested that clumps in molecular clouds are transient (Girart et al.
\cite{Girart02}; Viti et al. \cite{Viti03}).  Fig.~\ref{fmass} also shows that
the mass of the clumps becomes closer to the virial mass when they get bigger
and more massive, which is when they have a higher chance to form stars.
However, the mass calculated for the clumps is derived assuming constant CS
abundance.  If, as it seems, the CS abundance is lower for more evolved
clumps, the slope of the relation shown in Fig.~\ref{fmass} would be steeper,
and the clumps would probably have a mass closer to the virial mass.

\subsubsection{Largest clump and N$_{2}$H$^+$ clumps properties}

Figure~\ref{chancomb} shows the comparison of two channel maps for the
combined CS \J{1}{2} and N$_2$H$^+$ \J{1}{0} emission. We can clearly see that
the peak of the CS emission is shifted with respect to the peak of N$_2$H$^+$.
Moreover, the CS emission is more spread than N$_2$H$^+$, so we are detecting
chemical differentiation inside the clump.

The largest CS clump is not only the clump with the highest $M_{\rm
  clump}/M_{\rm virial}$ ratio (and thus with the highest probability of
undergoing star formation) but it is also the only one associated with
N$_{2}$H$^+$ emission (see Sect.~\ref{Sclump}).  This confirms the Taylor et
al.  (\cite{Taylor96}, \cite{Taylor98}) scenario. Interestingly, the density
of this clump, as derived from the CS emission, is similar to that of the
other clumps.  This may be explained if the emission of the N$_{2}$H$^+$
traces the highest densities of the clump, high enough for CS depletion to
occur, as has been observed in several dense starless cores (\eg\ Tafalla \et\
\cite{Tafalla02}, \cite{Tafalla04}).

We have calculated from the BIMA data the mass traced by the N$_2$H$^+$ for
this clump. Assuming a N$_2$H$^+$ fractional abundance of $8\times10^{-11}$,
which was obtained using the H$_2$ column density determination from the submm
sources of Visser \et\ (\cite{Visser02}), we could derive a total mass for the
biggest clump of 1.2~\mo. This mass is significantly higher than those of the
other clumps. In addition, $M_{\rm clump}/M_{\rm vir}$ is 0.52, so it is close
to collapse.

%

\subsubsection{On the small scale structure of L673}

The previous results, along with the results from the lower-resolution
analysis in Sect.~\ref{columndensities}, show at least the differences between
the two regions within the BIMA observations. The northern region shows a gas
less chemically evolved, and smaller and more numerous clumps. The southern
region contains the more evolved gas and the biggest clump.

The clumps detected in our observations seem to be of a different kind than
the ones usually studied in surveys of ``starless cores'' (Tafalla et al.
\cite{Tafalla02}; Bergin et al. \cite{Bergin01}). Those are usually older
cores, which have already achieved the more evolved and more massive stage of
evolution, as is clearly indicated by the detection of late-time molecules, as
N$_2$H$^+$ and NH$_3$, and by the high level of depletion of some molecules
onto dust grains. We find one such clump in our region, which might be just
arriving at that evolved phase. This clump is at the same time the most
massive, shows emission of N$_2$H$^+$, and different emission peaks for CS and
N$_2$H$^+$, which could point to depletion. However, we detect many other
clumps, less massive, that show almost no emission of late-time molecules, and
which probably will disperse before they can achieve the ``starless core''
stage.

We face the question of what the origin of these transient clumps could be.
They are probably caused by interstellar turbulence (see e.g. Padoan
\cite{Padoan95}; Ballesteros-Paredes \et\ \cite{Ballesteros99}). In support of
this idea, we observe that the derived masses for our clumps are around
0.1\mo, which is also the mass-scale of turbulent fragmentation in the ISM
(Padoan \cite{Padoan95}). This small scale turbulence is probably related to
large-scale flows in the diffuse Galactic interstellar medium, which several
studies (Hartmann \et\ \cite{Hartmann01}; Bergin \et\ \cite{Bergin04}) propose
as a mechanism that can lead to the fast formation and dispersal of molecular
clouds, which could explain the apparent rapid star formation and short cloud
lifetimes observed in the regions of the solar neighborhood. On a smaller
scale, several authors postulate the possible existence of transient clumps.
Falle \& Hartquist (\cite{Falle02}) propose that slow-mode MHD waves could
create, under suitable conditions, large density enhancements, of the order of
a factor of 30, which could last on the order of 1~Myr, before dispersing back
to the original density. At the same time, Vázquez-Semadeni \et\
(\cite{Vazquez-ph}, \cite{Vazquez03}) and Ballesteros-Paredes \et\
(\cite{Ballesteros03}) also argue that cores within molecular clouds may not
be in hydrostatical equilibrium, and would be transient features generated by
the dynamical flow in the cloud, not necessarily requiring strong magnetic
fields.  These cores must then either collapse, if they have an excess of
gravitational energy, or re-expand and merge back into the surrounding
molecular cloud. The re-expansion time is expected to be longer than the
compression time because of the retarding action of self-gravity, and is
estimated to be of the order of a few free-fall times.  This would be
consistent with the result that typically there are more starless than
star-forming cores in molecular clouds and that, as in our case, most of the
cores do not appear to be gravitationally bound. Only sufficiently massive
cores would eventually undergo local collapse.

%
%

In any case, such models would have consequences for the physical and chemical
evolution of these regions (Williams \& Viti \cite{Williams02}).  Firstly, the
chemistry would be mainly ``young'', as the molecules would be destroyed when
the gas returns to a more diffuse state. Secondly, this mechanism would halt
the process of freeze-out of molecules onto dust, as the frozen-out molecules
would also return to the gas phase as the density of the gas goes back to the
diffuse state. Thus, these regions would not be completely depleted of
molecules in the gas phase, as is apparently shown by some observations (Gibb
\& Little \cite{Gibb98}).  Finally, the mass distribution of these transient
clumps probably determines the low-mass star formation rate.

However, an alternative possibility is that the more massive clumps were
formed by aggregation of smaller clumps into a massive one, which could then
be closer to collapse.  The analysis of our clumps allows us to make some
rough estimates to see if this can happen since transient clumps are expected
to survive for $\sim 1$~Myr (Falle \& Harquist 2002; V\'azquez-Semadeni et al.
2004). Two limiting cases can be studied. One is that of a 1-dimensional
structure, that is, the clumps are assumed to move inside a filamentary
structure, with relative velocities (between clumps) of $0.3$~km~s$^{-1}$.
This is the typical difference in the radial velocities of the clumps, which
should roughly give a reasonable lower limit for the true relative velocities.
In the course of $\sim 1$~Myr any of the observed clumps would have travelled
$\sim 63000$~AU or $\sim 3\farcm5$, which is the size of the field of view of
our maps.  The other case is that of a 3-dimensional (3-D) structure with
similar scales in the three dimensions.  In this case the projected area of
this volume in any particular direction is $A_{\rm proj} \simeq V_{\rm
  tot}^{2/3}$, where $A_{\rm proj}$ and $V_{\rm tot}$ are the area and volume,
respectively.  Let us assume that in a volume $V_{\rm tot}$ there are $n$
identical spherical clumps of radius $R$ distributed randomly, with a random
velocity distribution with an average velocity $\langle v_{\rm 3D}\rangle$.
The 3-D filling factor of the clumps for this volume is $f_{\rm 3D}=n V_{\rm
  cl}/V_{\rm tot}= n V_{\rm cl}/A_{\rm proj}^{3/2}$, where $V_{\rm cl}$ is the
clump volume.  The volume swept up by a clump in a time $t$ is $V_{\rm
  sw}(t)=v_{\rm 3D}\,t\,\pi R^2$.  For a given clump, the time scale needed to
cross another clump, $t_{X}$, is such that $f_{\rm 3D}\,V_{\rm
  sw}(t_{X})=V_{\rm cl}$.  From the previous equations, the average crossing
scale time can be written as $\langle t_{X}\rangle=A_{\rm
  proj}^{3/2}/(n\,\langle v_{\rm 3D}\rangle\,\pi R^2)$.  Since what we measure
is the radial velocity, $v_{\rm obs}$ and $\langle v_{\rm
  3D}\rangle=2\,\langle v_{\rm obs}\rangle$, then $\langle t_{X}\rangle=A_{\rm
  proj}^{3/2}/(2\,n\,\langle v_{\rm obs}\rangle\,\pi R^2)$.  For $A_{\rm
  proj}$ equal to our field of view ($\sim3\farcm5\times2\farcm0 $), which
includes 15 clumps, and using a radius equal to the average value for the
measured clumps (from Table~7), $R\simeq18''$, we found that $\langle
t_{X}\rangle\simeq6\times10^5$~yr.  Then it could be possible that star
formation was the result not of the gravitational collapse of just one clump,
but of a ``merging'' effect. In this case, the geometry of dense cores would
be far from spherical symmetry, even the density distribution would not follow
$n(r) \propto r^{-2}$. Tafalla et al. (\cite{Tafalla04}) find that their
starless cores are not spherical, but that clear inhomogeneities are found in
them.

\section{Conclusions}

We have made a multitransitional study of the spatial distribution of the
molecular emission with the 14 m FCRAO telescope in the starless core found in
the L673 region, which had been previously observed at lower angular
resolution (Morata \et\ \cite{Morata97}) and with the BIMA array (MGE03). The
main goal of these observations was to combine them with the array
observations in order to obtain high resolution maps of the full emission, and
thus be able to confirm that the clumpy structure detected with the BIMA
telescope was real.  The main results were:

%
     \begin{figure} [t]
      \includegraphics[angle=-90,width=\hsize]{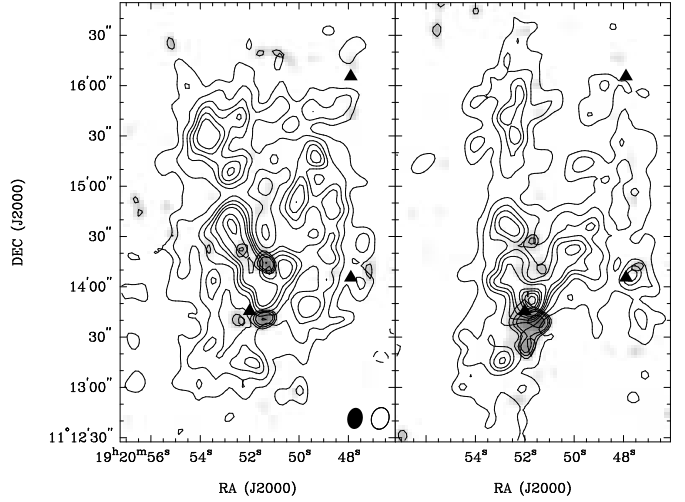}
     \caption[]{ Comparison of the channel maps of the combined CS \J{2}{1}
      emission (solid contours) for the 7.16 ({\it left}) and 6.87 km s$^{-1}$
      ({\it right}) channels; and the BIMA N$_2$H$^+$ \J{1}{0} emission
      (contours and grey scale) for the $-0.8$ ({\it left}) and $-1.1$ km
      s$^{-1}$ ({\it right}) channels. The contour levels for the CS maps are
      the same as in Fig.~\ref{fchannel}. The contour levels for the
      N$_2$H$^+$ maps are 3, 3.5, 4, 5, 6, and 7 times 0.20~Jy~beam$^{-1}$
      (the rms noise of the maps). The symbols are the same as those in
      Fig.~\ref{ffcrao}.  }
     \label{chancomb} 
     \end{figure}
%

\begin{enumerate}

\item With the FCRAO we detected emission in the CS \J{2}{1}, C$^{34}$S
  \J{2}{1}, HCO$^+$ \J{1}{0}, and H$^{13}$CO$^+$ \J{1}{0} lines. Although
  every map shows clear morphological differences, we found that the high
  density molecular emission in the four molecules arises from a filamentary
  structure oriented in the NW-SE direction. The ratio between HCO$^+$ and
  H$^{13}$CO$^+$ also confirms that the apparent lack of HCO$^+$ \J{1}{0}
  emission in the BIMA maps was due to the extremely high optical depth in
  this line. We also found that several submm sources detected in this region
  are located in strong molecular line emitting regions, and that they seem to
  be well correlated with the relative intensity peaks of H$^{13}$CO$^+$.

\item We selected several positions to study the local physical parameters. We
  confirm the results of MGE03: the column density of CS is higher than that
  of HCO$^+$ for positions were the gas is chemically younger, i.e. not
  associated with IRAS or submm sources. In the positions where there is a
  submm or IRAS source, the HCO$^+$ column density is higher than that of CS.

\item The BIMA and FCRAO combined maps of the CS emission show that all the
  prominent clumps found in the BIMA observations are also detected but with
  more intensity. We believe that these clumps are real. We also recover
  emission from other structure which was previously undetected or only
  marginally detected. The overall aspect is also of a filamentary structure
  connecting several intense clumps.

\item We found a total of 15 clumps in our combined data cube.  All of them
  are resolved at our angular resolution, with diameters in the 0.03$-$0.09~pc
  range. Their estimated masses range between 0.02 and 0.36 \mo, and their
  densities between 0.6 and 4.6$\times10^{4}$\cmt, which indicates that the CS
  \J{2}{1} emission is clearly subthermalized. 

\item There is a clear segregation in the properties of the detected clumps
  between the northern and southern regions of the map. The northern region
  shows the less chemically evolved gas, and less massive but more numerous
  clumps. On the other hand, the southern region contains the more evolved gas
  and the most massive clump. This clump is also the only one associated with
  BIMA N$_2$H$^+$ emission. Moreover the H$^{13}$CO$^+$ emission correlates
  well with the dust emission traced by the submillimeter observations, and
  its emission peak also coincides with the BIMA N$_2$H$^+$ peak and the SMM 5
  source. All these results support the predictions of our chemical model,
  which proposed that late-time molecules, such as N$_2$H$^+$ and HCO$^+$,
  should be found in the same regions.

\item The clump masses derived are below the virial mass, which means that
  these clumps must be transient, as only the more massive ones are able to
  condense into stars.

\item The properties of the clumps we detect seem to be different from those
  of the ``starless cores'' recently found in the literature. Although we
  might have detected one such core, the rest are in an early evolutionary
  stage, after which they will most likely disperse.

\end{enumerate}

We find that the simple chemical model proposed in MGE03 to explain the
properties of the CS core of L673 is confirmed by the combination of
single-dish and array observations. However, there is still a lot of work to
be done on the formation mechanisms of these clumps and on the possibility
that the merging of smaller clumps can lead to more massive and
gravitationally unstable clumps. Further observing and modelling of the
chemistry of these clumps and of the interclump gas will be needed in order to
test the transient nature of the clumps.

\begin{acknowledgements}

  OM acknowledges the support of the Na\-tional Science Foundation to the
  astrochemistry group at Ohio State Uni\-versity. JMG and RE are supported by
  SEUI (Spain) grant AYA2002-00205. FCRAO is supported by NSF grant
  AST~02-28993. We thank Mark Heyer for carrying out the observations and for
  the helpful support in preparing the observations. This publication makes
  use of data products from the Two Micron All Sky Survey, which is a joint
  project of the University of Massachusetts and the Infrared Processing and
  Analysis Center/California Institute of Technology, funded by the National
  Aeronautics and Space Admi\-nistration and the National Science Foundation.

\end{acknowledgements}



\begin{thebibliography}{}
\bibitem[2003]{Ballesteros03} Ballesteros-Paredes, J., Klessen, R.~S., \&
  V\'{a}zquez-Semadeni, E.\ 2003, ApJ, 592, 188
\bibitem[1999]{Ballesteros99} Ballesteros-Paredes, J., V\'{a}zquez-Semadeni, \&
  E., and Scalo, J.\ 1999, ApJ, 515, 286
\bibitem[2001]{Bergin01} Bergin, E.~A., Ciardi, D.~R., Lada, C.~J., Alves, J.,
  \& Lada, E.~A. 2001, ApJ, 557, 209
\bibitem[2004]{Bergin04} Bergin, E.~A., Hartmann, L.~W., Raymod, J.~C.,
  Ballesteros-Paredes, J. 2004, ApJ, 612, 921
\bibitem[2002]{Falle02}
 Falle, S.~A.~E.~G., \& Hartquist, T.~W.\ 2002, MNRAS, 329, 195 
\bibitem[1998]{Gibb98} Gibb, A.~G., \& Little, L.~T. 1998, MNRAS, 295,299
\bibitem[1999]{Girart99} 
 Girart, J.~M., Ho, P.~T.~P., Rudolph, A.~L., Estalella, R., Wilner, D.~J., 
 \& Chernin, L.~M. 1999, ApJ, 522, 921
\bibitem[2000]{Girart00} 
 Girart, J.~M., Estalella, R, Ho, P.~T.~P., \& Rudolph, A.~L. 2000, ApJ, 
 539, 763
\bibitem[2002]{Girart02} 
 Girart, J.~M., Viti, S., Williams, D.~A., Estalella, R, \& Ho, P.~T.~P. 2002,
 A\&A, 388, 1004
\bibitem[2001]{Hartmann01}Hartmann, L.~W., Ballesteros-Paredes, J., Bergin,
 E.~A. 2001, ApJ, 562, 852
\bibitem[2003]{Helfer03}
 Helfer, T.~T., Thornley, M.~D., Regan, M.~W., \et\ 2003, ApJS, 145, 259 
\bibitem[1983]{Herbig83} Herbig, G.~H., \& Jones, B.~F. 1983, AJ, 88, 1040
\bibitem[2000]{Johnstone00}Johnstone, D., Wilson, C.~D., Moriarty-Schieven,
  G., \et\ 2000, ApJ, 545, 327
\bibitem[1978]{Langer78}
 Langer, W.~D., Wilson, R.~W., Henry, P.~S., \& Gu\'{e}lin, M. 1978, ApJ, 225, 
 L139
\bibitem[1993]{Langer93} Langer, W.~D., \& Penzias, A. 1993,
  ApJ, 408, 539
\bibitem[1997]{Morata97}Morata, O., Estalella, R., L\'{o}pez, R., \& Planesas,
P. 1997, MNRAS, 292, 120
\bibitem[2003]{Morata03}  Morata, O., Girart, J.~M., \& Estalella, R. 2003,
  A\&A, 397, 181 (MGE03)
\bibitem[1998]{Motte98} Motte, F., Andr\'{e}, P., \& Neri R. 1998, A\&A,
  336,150
\bibitem[1991]{myers}Myers, P.~C., Fuller, G.~A., Goodman, A.~A., \& Benson,
P.~J. 1991, ApJ, 376, 561
\bibitem[1995]{Padoan95} Padoan, P.\ 1995, MNRAS, 277, 377
\bibitem[1991]{Pastor91}Pastor, J., Estalella, R., L\'opez, R., et al. 1991,
  A\&A, 252, 320
\bibitem[1999]{Stanimirovic99}
 Stanimirovic, S., Staveley-Smith, L., Dickey, J.~M., Sault, R.~J., \&
 Snowden, S. 1999, MNRAS, 302, 417
\bibitem[2004]{Tafalla04} 
 Tafalla, M., Myers, P.~C., Caselli, P., \& Walmsley, C.~M.\ 2004, \aap, 
416, 191 
\bibitem[2002]{Tafalla02}
 Tafalla, M., Myers, P.~C., Caselli, P., Walmsley, C.~M., \& Comito, C.\ 2002, 
 ApJ, 569, 815 
\bibitem[1996]{Taylor96}
 Taylor, S.~D., Morata, O., \& Williams, D.~A.\ 1996, A\&A, 313, 269 
\bibitem[1998]{Taylor98}
 Taylor, S.~D., Morata, O., \& Williams, D.~A.\ 1998, A\&A, 336, 309 
\bibitem[2003]{Vazquez03} V\'{a}zquez-Semadeni, E., Ballesteros-Paredes, J., \&
  Klessen, R.~S.\ 2003, ApJ, 585, L131
\bibitem[2005]{Vazquez-ph} V\'{a}zquez-Semadeni, E., Kim, J., Shadmehri M., \&
  Ballesteros-Paredes, J. 2005, ApJ, 618, 344
\bibitem[2002]{Visser02}
 Visser, A.~E., Richer, J.~S., \& Chandler, C.~J. 2002, AJ, 124, 2756
\bibitem[2003]{Viti03}
 Viti, S., Girart, J. M., Garrod, R., Williams, D.~A., \& Estalella, R. 2003, 
  A\&A, 399, 187
\bibitem[1984]{Vogel84}
 Vogel, S.~N., Wright, M.~C.~H., Plambeck, R.~L., \& Welch, W.~J. 1984, 
 ApJ, 283, 655
\bibitem[2002]{Williams02} Williams, D.~A., \& Viti, S., 2002, in Chemistry as
  a Diagnostic of Star Formation, ed.\ C.~L. Curry, \& M. Fich (Ottawa, Canada:
  NRC Press), 106
\bibitem[1994]{Williams94}
 Williams, J.~P., de Geus, E.~J., \& Blitz, L. 1994, ApJ, 428, 693
\bibitem[1991]{Ye91} Ye, T., Turtle, A.~J., \& Kennicutt, R.~C.\ 1991, MNRAS,
  249, 722
\bibitem[1989]{zhou}Zhou, S., Wu, Y., Evans, N.~ J., Fuller, G.~ A., \& Myers,
P. C. 1989, ApJ, 346, 168

\end{thebibliography}
\end{document}